\documentclass[preprint,showpacs,amsmath,amssymb,pre,aps,superscriptaddress]{revtex4-1}
\usepackage{graphicx}
\usepackage[tight,TABTOPCAP]{subfigure}
\usepackage[]{SIunits}
\usepackage[usenames, dvipsnames]{color}
\usepackage{natbib}
\usepackage[normalem]{ulem}

\usepackage[utf8]{inputenc}
\usepackage{comment}
\usepackage[bordercolor=white,backgroundcolor=gray!30,linecolor=black,colorinlistoftodos]{todonotes}

\begin{document}

\title{Open-Boundary Hamiltonian adaptive resolution. From grand canonical to non-equilibrium molecular dynamics simulations}

\author{Maziar Heidari}
\affiliation{Max Planck Institute for Polymer Research, Ackermannweg 10, 55128 Mainz, Germany}

\author{Kurt Kremer}
\affiliation{Max Planck Institute for Polymer Research, Ackermannweg 10, 55128 Mainz, Germany}

\author{Ramin Golestanian}
\affiliation{Max Planck Institute for Dynamics and Self-Organization (MPIDS), 37077 Göttingen, Germany}
\affiliation{Rudolf Peierls Centre for Theoretical Physics, University of Oxford, Oxford OX1 3PU, United Kingdom}

\author{Raffaello Potestio}
\thanks{These authors contributed equally to this work.}
\email{raffaello.potestio@unitn.it}
\affiliation{Physics Department, University of Trento, via Sommarive, 14 I-38123 Trento, Italy}
\affiliation{INFN-TIFPA, Trento Institute for Fundamental Physics and Applications, I-38123 Trento, Italy}

\author{Robinson Cortes-Huerto}
\thanks{These authors contributed equally to this work.}
\email{corteshu@mpip-mainz.mpg.de}
\affiliation{Max Planck Institute for Polymer Research, Ackermannweg 10, 55128 Mainz, Germany}

\date{\today}
\begin{abstract}
 We propose an open-boundary molecular dynamics method in which an atomistic system is in contact with an infinite particle reservoir at constant temperature, volume and chemical potential. In practice, following the Hamiltonian adaptive resolution strategy, the system is partitioned into a domain of interest and a reservoir of non-interacting, ideal gas, particles. An external potential, applied only in the interfacial region, balances the excess chemical potential of the system. To ensure that the size of the reservoir is infinite, we introduce a particle insertion/deletion algorithm to control the density in the ideal gas region. We show that it is possible to study non-equilibrium phenomena with this open-boundary molecular dynamics method. To this aim, we consider a prototypical confined liquid under the influence of an external constant density gradient. The resulting pressure-driven flow across the atomistic system exhibits a velocity profile consistent with the corresponding solution of the Navier-Stokes equation. This method conserves, on average, linear momentum and closely resembles experimental conditions. Moreover, it can be used to study various direct and indirect out-of-equilibrium conditions in complex molecular systems.
\end{abstract}
\maketitle
\makeatletter
\let\toc@pre\relax
\let\toc@post\relax
\let\toc@pre\relax
\makeatother

\setlength{\parindent}{0pt}

\section{Introduction}

Computational and experimental communities routinely cooperate by comparing the results obtained from their respective methods. However, such comparisons are intrinsically limited in scope because real systems approach the thermodynamic limit, whereas model systems usually have a finite number of particles. Indeed, standard computer simulations frequently consider closed systems whose fixed number of particles introduces finite-size effects due to the {\it de facto} simultaneous use of different statistical ensembles \cite{cervera2011,Roman2008,Binder-etal-JPhysCondensMatter2-7009-1990}.

With the computational power nowadays available, it is tempting to ask whether it is possible to increase the size of the system and safely ignored ensemble finite-size effects, i.e. reach the thermodynamic limit.  In particular, for a system of total number of particles $N_{\rm 0}$ at temperature $T$ in a volume $V_{0}$, it is possible to consider a subdomain of volume $V$ with an average number of particles $\langle N\rangle$. The system is in the grand canonical ensemble if $V$ is of the order of 1$\%$ of the total volume $V_{\rm 0}$ \cite{Heidari2018Frenkel}. This size constraint implies using huge simulation boxes which, in most cases, demand a tremendous computational effort.

A simplification of the physical representation of the particles in the reservoir alleviates this computational load. This idea is the essence of the adaptive resolution method: an atomistic sub-domain of volume $V$, defined within the simulation box, is in contact with a reservoir of coarse-grained particles \cite{adress1,adress2,adress3,annurev,adresstoluene}. A smooth interpolation between atomistic and coarse-grained forces, acting on molecules present at the interface between the two regions, ensures a consistent description of the whole system. Indeed, the adaptive resolution framework is a robust method to perform simulations in the grand canonical ensemble \cite{DelleSite-Grand,Agarwal_2015,dellesite_GC2019}. To build upon these results, we discuss two components that one should consider to make use of the adaptive resolution method to perform well-controlled open-boundary molecular dynamics simulations. 

First, a simulation in the $\mu VT$ ensemble requires to know beforehand the chemical potential $\mu$ of the bath. This condition is analogous to the situation in the  $NVT$ ensemble, in which one first requires to fix the number of particles $N$. In the original adaptive resolution framework, it is possible to obtain the difference in chemical potential between atomistic and coarse-grained representations of the system \cite{DelleSiteChemPot}. If one knows the chemical potential of the coarse-grained model, then it is straightforward to obtain and control $\mu$ for the atomistic one. It is thus convenient to use the simplest possible coarse-grained representation such that the calculation of the chemical potential does not involve an additional external computation.  

Second, a system is in the grand canonical ensemble if it is in thermal and chemical equilibrium with an infinite particle reservoir.  Computer memory limitations prevent the possibility to consider an infinite number of particles. Instead,  a particle insertion/deletion algorithm coupled to the simulation setup effectively ensures this requirement by allowing the interchange of particles with an infinite ideal gas reservoir. In the context of the adaptive resolution method, a semi-grand canonical method was proposed to illustrate this point \cite{Mukherji2013}. The idea to interpret the AdResS methodology in terms of a simulation with a particle reservoir was first formulated in Ref.\ \cite{GCAdResS} and subsequently generalised to atomistic/mesoscopic continuum adaptive
resolution models \cite{AdResS-continuum2009,ALEKSEEVA201614} in which open boundary conditions can be readily enforced \cite{Delgado-Buscalioni2015,DELLESITE20171,ZAVADLAV20182352}.
 Nevertheless, the sampling resulting from particle insertion/deletion events, even if only applied in the coarse-grained region, become less representative as the density, concentration and complexity of the system increases.

To consider these points, here we present a method that combines a particle insertion/deletion algorithm with the Hamiltonian adaptive resolution framework ({\tt H-AdResS}) \cite{hadress,MC_hadress}. Following the method suggested in Ref. \cite{SPARTIAN}, we replace the coarse-grained model by a reservoir of non-interacting thermalised particles (ideal gas).  In this case, the applied external potential used to ensure a uniform density through the simulation box balances the excess chemical potential of the atomistic model. Therefore, the atomistic system is at constant chemical potential with a reservoir of ideal gas particles. We introduce the particle insertion/deletion algorithm, operating on the ideal gas reservoir, to overcome the limitations existing on available methods due to high density/concentration and system complexity conditions. The method is thus capable of performing constant $\mu$ molecular dynamics simulations without the necessity to include external forces and to compensate for depletion of particles in the reservoir \cite{Perego2016,Claudio2018}.

It is possible to study non-equilibrium phenomena with this open-boundary molecular dynamics method. In particular, we consider a confined liquid such that its ideal gas reservoir is under the influence of a constant density gradient.  Initially, a uniform density profile is enforced parallel to the surfaces (see Figure \ref{Fig:Model}). Upon equilibration, a density gradient imposed in the reservoir induces a pressure-driven flow in the system with a velocity profile consistent with the corresponding solution of the Navier-Stokes equation. The method closely resembles experimental conditions and avoids artefacts present in current methods due to the non-conservation of linear momentum.  Furthermore, the intrinsic arbitrariness of the ideal gas reservoir opens the possibility to study various direct and indirect non-equilibrium conditions.

The paper reads as follows: In section \ref{model}, we validate the use of {\tt H-AdResS} to study confined liquids and introduce the particle insertion/deletion algorithm. We present the results for pressure-driven flow in section \ref{poiseuille} and finally discuss, conclude and outline research directions in section \ref{conclusions}.

\section{Model}\label{model}

In adaptive resolution simulations it is possible to couple a target system with an ideal gas reservoir of particles at constant chemical potential \cite{kreis_EPJST2015,SPARTIAN}. Particularly, it is possible to set the chemical potential of the target system by controlling the number density in the ideal gas reservoir. In this work, we implement a particle insertion algorithm that operates on the ideal gas region and permits fluctuations in the number density around a target value. Consequently, the standard {\tt H-AdResS} setup now allows one to perform open-system molecular dynamics simulations in equilibrium and, more important, nonequilibrium conditions. As an illustration, we study the prototypical example of liquid flow across a narrow channel. 

Before introducing the particle insertion algorithm, we validate {\tt H-AdResS} to study a confined liquid between parallel walls.  

\subsection{{\tt H-AdResS} for a confined liquid}

 \begin{figure}[ht]
 \begin{center}
  \includegraphics[width=0.5\textwidth,angle=0]{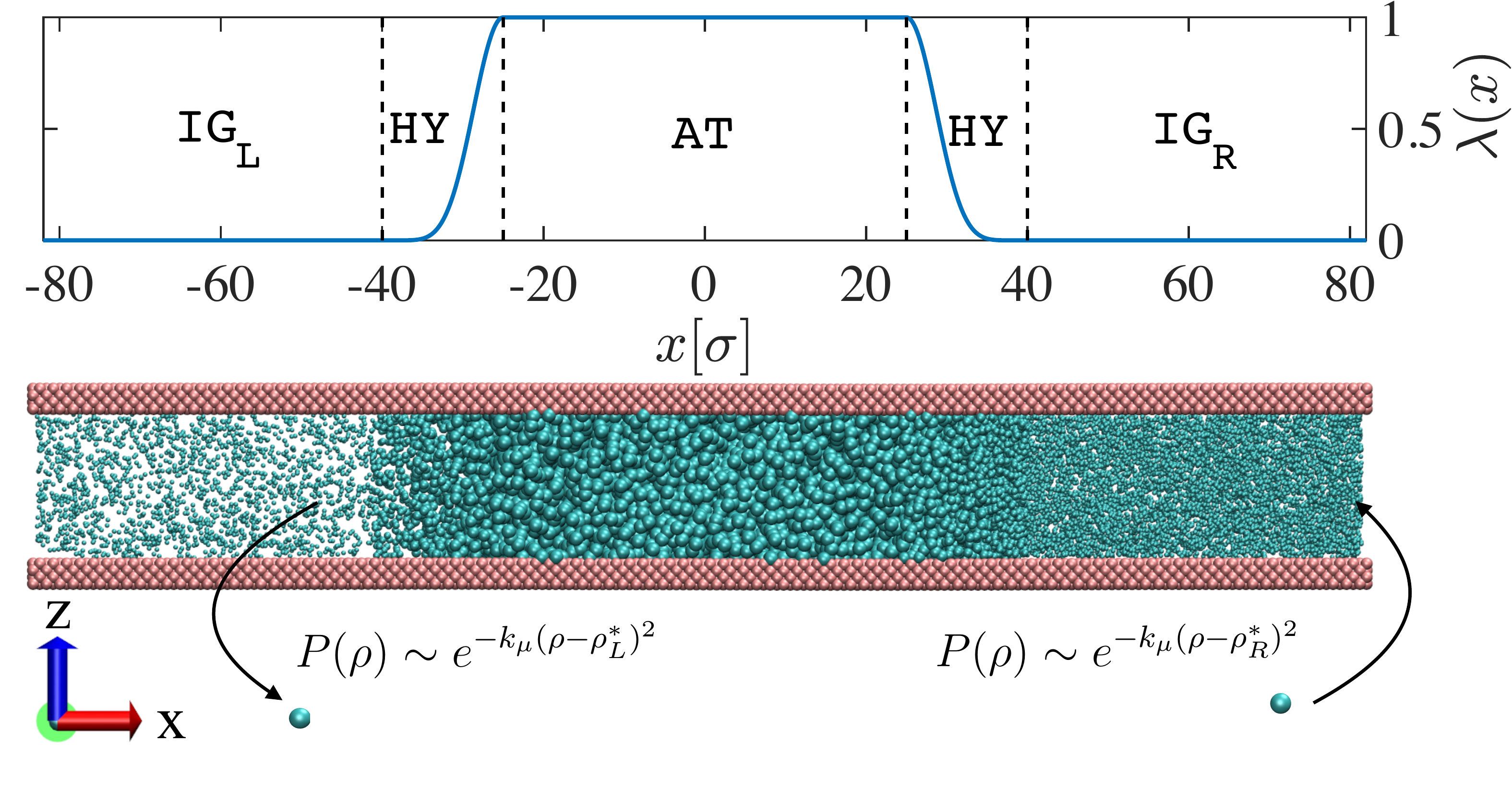}
 \caption{ 
 Hamiltonian adaptive resolution setup used to simulate a liquid confined between parallel plates. Upper panel: the simulation box is partitioned into atomistic (AT), hybrid (HY) and two unconnected left and right reservoirs of ideal gas (IG$_{\rm L/R}$) particles. The identity of the molecules is defined by the switching function $\lambda(x)$ that takes values between 0 and 1. Lower panel: simulation snapshot explicitly showing the confined liquid in the various subdomains. For this particular example, the number density is lower in the IG$_{\rm L}$ than in the IG$_{\rm R}$ region.} 
 \label{Fig:Model}
 \end{center}
 \end{figure}
 
Let us consider a single component liquid confined between parallel walls with normal perpendicular to the $x$-axis and separated by a distance $D$. In the adaptive resolution method (AdResS) \cite{adress1,adress2,adress3}, atomistic (AT) and coarse grained (CG) representations of the system are concurrently present in the simulation box and coexist in thermodynamic equilibrium \cite{adress1,adress2,adress3}.  In particular, by using a position-dependent switching field  $\lambda(x)$ (see Figure \ref{Fig:Model}) it is possible to write a Hamiltonian ({\tt H-AdResS}) \cite{hadress,MC_hadress} for the whole system. Since AA and CG representations of the system follow different equations of state, an additional term is included in the Hamiltonian to guarantee a constant density (implying constant chemical potential) across the simulation box.  Finally, the choice of the coarse grained representation is rather arbitrary and can even be reduced to a reservoir of noninteracting - ideal gas (IG) - particles \cite{kreis_EPJST2015,SPARTIAN}. 
In this context we write the Hamiltonian for a system composed of $N=N_{\rm s}+N_{\rm w}$ \emph{solvent} (s) and \emph{wall} (w) molecules as \cite{Espanol-etal-JCP142-064115-2015}:
\begin{equation}\label{eq:ham}
H_{[\lambda]}(r,p) = K + V^{\rm w}(r) + V_{[\lambda]}(r)\, ,
\end{equation}
with $(r,p)$ positions and momenta and $K=\sum_{i=1}^{N}\mathbf{p}_{i}^{2}/2m_{i}$ the total kinetic energy of the system. The term
\begin{equation*}
V^{\rm w}(r)=\frac{\kappa}{2}\underset{j,j'\in\rm w}{\sum_{j\neq j'}^{N_{\rm w}}}(r_{j,j'}-r_{0})^{2}\, ,
\end{equation*}
is a harmonic, nearest-neighbour $V^{\rm w} (r>r_{\rm cut})=0$ potential used to restraint the position of the wall molecules. $\kappa$ is the spring constant, $r_{0}$ is the equilibrium length and $r_{j,j'}$ is the distance between a given pair of wall particles. The wall particles are located on a fcc lattice of density $0.9\, \sigma^{-3}$ with parameters $\kappa=1000\, \epsilon\sigma^{-2}$ and $r_0=1.1626\, \sigma$. $V^{\rm w}(r)$ is only applied during the initial equilibration. For production runs, the surface particles are frozen in their final equilibrium positions. 

The last term in the Hamiltonian \eqref{eq:ham} is written as:  
\begin{align}\label{eq:vlam}
V_{[\lambda]}&(r) = \underset{i \in\rm s}{\sum_{i=1}^{N_{\rm s}}}\left[ \lambda_{i}\left\{\frac{1}{2}\underset{i' \neq i}{\sum_{i'=1}^{N_{\rm s}}}U^{\rm s,s}(r_{i,i'})+\underset{j \in\rm w}{\sum_{j=1}^{N_{\rm w}}}U^{\rm s,w}(r_{i,j})\right\}+(1-\lambda_{i})\underset{j \in\rm w}{\sum_{j=1}^{N_{\rm w}}}U^{\rm ig,w}(r_{i,j}) - \Delta H^{\rm s}(\lambda_{i})\right]\, .
\end{align}
The solvent molecules $i$ interact via an intermolecular potential modulated by a switching field $\lambda_{i}\equiv \lambda(x_{i})$ (see Figure \ref{Fig:Model}). The switching field takes values 0 in the IG region where solvent particles only interact repulsively with wall particles and 1 in the AT region where solvent particles fully interact with both solvent and wall particles. A smooth interpolation between 0 and 1 is defined in the hybrid (HY) region. The free energy compensation (FEC) term, $\Delta H^{\rm s}$, compensates non-physical forces due to gradients of the switching field and guarantees a uniform density throughout the simulation box. Note that the FEC is only calculated for the confined liquid particles. The starting point of the simulations corresponds to a fully atomistic case. Once these simulations are equilibrated, the adaptive resolution setup is turned on and $\Delta H^{\rm s}$ is computed \emph{on-the-fly} following the procedure described in Ref.\ \cite{Heidari2016} included in the equilibration of the {\tt H-AdResS} run. Finally, for homogeneous systems, $\Delta H^{\rm s}(\lambda=1)$ corresponds to the system's excess chemical potential $\mu_{\rm exc}$. This procedure has been identified with a thermodynamic integration in space, i.\ e.\ spatially-resolved thermodynamic integration (SPARTIAN) \cite{SPARTIAN}. 

We model all the solvent-wall interactions with the truncated and shifted Lennard-Jones potential 

\begin{equation}
U^{\alpha,\beta}(r)=4\epsilon_{\alpha,\beta}\left[\left( \frac{\sigma_{\alpha,\beta}}{r} \right)^{12}-\left( \frac{\sigma_{\alpha,\beta}}{r} \right)^{6}\right]\, ,
    \label{eq:tslj}
\end{equation}

where the units of length, energy and mass are defined by the parameters $\sigma$, $\epsilon$ and $m$, respectively. In the following, we report the results in LJ units with time $\sigma(\epsilon/m)^{1/2}$, temperature $\epsilon/k_{\rm B}$ and pressure $\epsilon/\sigma^{3}$. For a given cutoff radius $r_{\rm cut}$ the value $U^{\alpha,\beta}(r_{\rm cut})$ is evaluated and subtracted from Eq.\ \eqref{eq:tslj}. The parameters used to describe all interactions between species ($\alpha,\, \beta$) for different regions within the simulation box are presented in Table \ref{table1} with fixed solvent-solvent (s,s) and ideal gas-wall (ig,w) values. For solvent-wall (s,w) interactions in the AT region we consider the purely-repulsive Lennard-Jones potential (WCA) as well as truncated and shifted Lennard-Jones potential with $r_{\rm cut}=2.5\sigma$ with varying interaction strength modulated by the parameter $\eta$ with $\eta=0.5$, 1.0, 1.5, 2.0 and 2.5. 

\begin{table}[h]
\begin{tabular}{| l | c | c | c |}
\hline
($\alpha$,$\beta$) & $\epsilon_{\alpha,\beta}$ & 
$\sigma_{\alpha,\beta}$ & $r^{\rm cut}_{\alpha,\beta}$ \\
\hline
\hline
(s,s) & $\epsilon$ & $\sigma$ & 2.5$\sigma$\\
\hline
(s,w) & $\eta \epsilon$ & $\sigma$ & 2.5$\sigma$\\
\hline
(ig,w) & $\epsilon$ & $\sigma$ & $2^{1/6}\sigma$\\
\hline
\end{tabular}
\caption{Lennard-Jones parameters used to describe solvent-solvent and solvent-wall interactions. Additionally to the purely repulsive case (not included in this table), the latter interactions are modulated in the AT region by the parameter $\eta$, whereas in the IG region is defined as purely repulsive.}\label{table1}
\end{table}

\begin{figure}[h]
 \begin{center}
  \includegraphics[width=0.45\textwidth,angle=0]{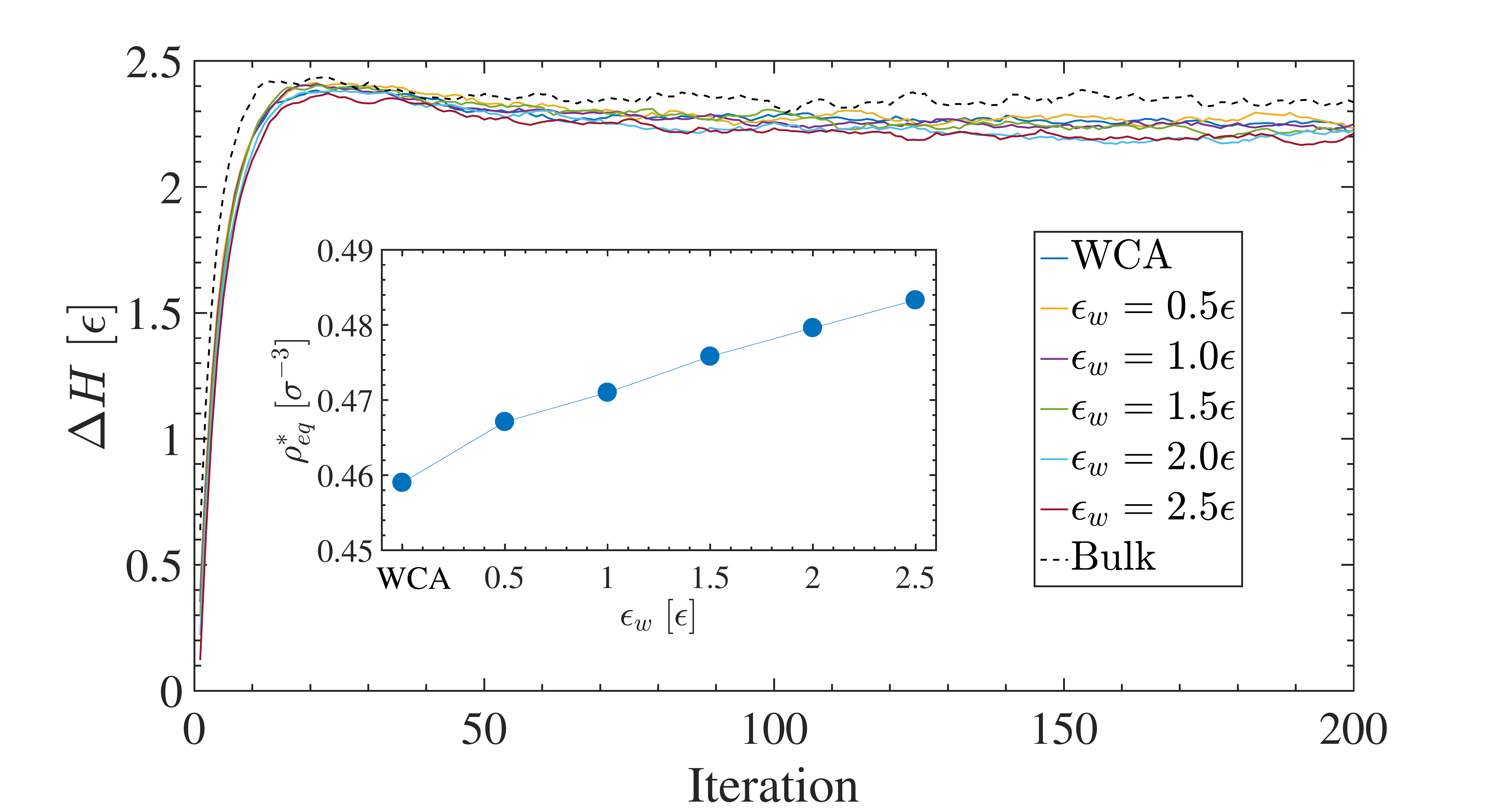}
 \caption{ The convergence of compensating energies of LJ fluid for different fluid-wall interactions as obtained by the method outlined in Ref.\ \onlinecite{SPARTIAN}. In all cases, the initial configuration is equilibrated at the same pressure $P=2.65\, \epsilon\sigma^{-3}$ and temperature $k_{\rm B}T=2.0\, \epsilon$. Inset shows the final equilibrium densities obtained from fully atomistic simulations of the systems.}
 \label{Fig:exchem_Single}
 \end{center}
 \end{figure}

In this example, the number of solvent and wall particles is fixed with $N_s=97020$ and $N_w=48000$, respectively. The size of the box is set by $L_x=164.41\, \sigma$ and $L_y=49.32\, \sigma$ while $L_z$ is fixed by the system's pressure with variations in the range of $L_z= 24.75\, \sigma,\cdots,25.61\, \sigma$. For the case of homogeneous liquid, i.e. no confining walls, $L_z=18.74\, \sigma$. The initial fully atomistic equilibration is carried out in the NPT ensemble using Nose-Hoover thermostat and barostat for $5000\, \tau$ with time step size of $\delta t=10^{-3}\, \tau$. The temperature is fixed at $k_BT=2.0\, \epsilon$ with damping coefficient $10\, \tau$ and the pressure is fixed anisotropically at $P=2.65\, \epsilon\sigma^{-3}$ with damping coefficient $100\, \tau$ by applying a force normal to the walls. The final equilibrium density, which is defined as the ratio of number of solvent particle to the total volume of the simulation box ($\rho^*_{eq}=N_S/V$), is reported in the inset of Figure \ref{Fig:exchem_Single} for different fluid-wall interaction.

To validate the reliability of {\tt H-AdResS} to study confined liquids, we verify that the FEC terms, $\Delta H^{\rm s}$, converge. The size of AT and HY regions is $50\, \sigma$ and $15\, \sigma$ and the {\tt H-AdResS} parameters are listed in Table \ref{SPARTIAN_Parameters}. As it was mentioned before, the application of the FEC on the system leads to a constant density profile across the simulation box. In this case, the system reaches an equilibrium state in which atomistic and ideal gas particles have equal chemical potential. The evolution of the FEC as a function of time is plotted in Figure \ref{Fig:exchem_Single}. After 200 iterations, which corresponds to $2000\, \tau$ simulation time, (see Table \ref{SPARTIAN_Parameters} for more detail), the algorithm converges to $\mu=2.33\pm0.008\, \epsilon$ for the bulk liquid (no confinement), in a good agreement with the previously reported value for LJ fluid ($\mu=2.33\pm0.01\, \epsilon$ \cite{SPARTIAN}). Somewhat expected, the obtained values of $\Delta H^{\rm s}$ are inversely proportional to the equilibrium density of the system for every liquid wall interaction case. 
\begin{table}[h]
\begin{center}
    \begin{tabular}{c c c c c c c c c c c c p{5cm} |} 
    \hline
     $\Delta t^{dr}_{smp}$ & $\Delta t^{th}_{smp}$ & $\Delta t^{dr}_{ave}$ & $\Delta t^{th}_{ave}$ & $\delta t$ & Iterations\\ \hline
       			100	 & 100 & $5\times10^4$ & $5\times10^4$ & $10^{-3}$ $\tau$ & 200 \\ \hline
    \end{tabular}
\end{center}
\begin{center}
    \begin{tabular}{c c c c c c p{5cm} |} 
    \hline
     $\Delta \lambda$ & $\Delta X [\sigma]$ & $c [\epsilon]$ & $\rho^* [\sigma^{-3}]$ & $\sigma_{\rho} [\sigma]$ & $l [\sigma]$\\ \hline
       			0.005	 & 0.5 & 1.0 & 0.5 & 2.0 & 4.0 \\ \hline
    \end{tabular}
\end{center}
\caption{{\tt H-AdResS} parameters for LJ and WCA fluids. The first two rows correspond to the parameters being used in time sampling of drift and thermodynamic forces ($t^{dr}_{smp}$ and $t^{th}_{smp}$) as well as their averaging process ($t^{dr}_{ave}$ and $t^{th}_{ave}$). The values are presented in number of time steps. The second two rows enlist the size of the bins in discretization of HY region to obtain drift force ($\Delta \lambda$) and thermodynamic force ($\Delta X$). The parameters being used in the convolution of the density profile with Gaussian function are the thermodynamic scaling factor ($c$), reference number density ($\rho^*$), the standard deviation of Gaussian function ($\sigma_\rho$) and the domain of convolution integration ($l$). Details about the precise meaning of these parameters can be found in Ref.\ \cite{SPARTIAN}.}\label{SPARTIAN_Parameters}
\end{table}

 \begin{figure*}[h]
 \begin{center}
  \includegraphics[width=0.9\textwidth,angle=0]{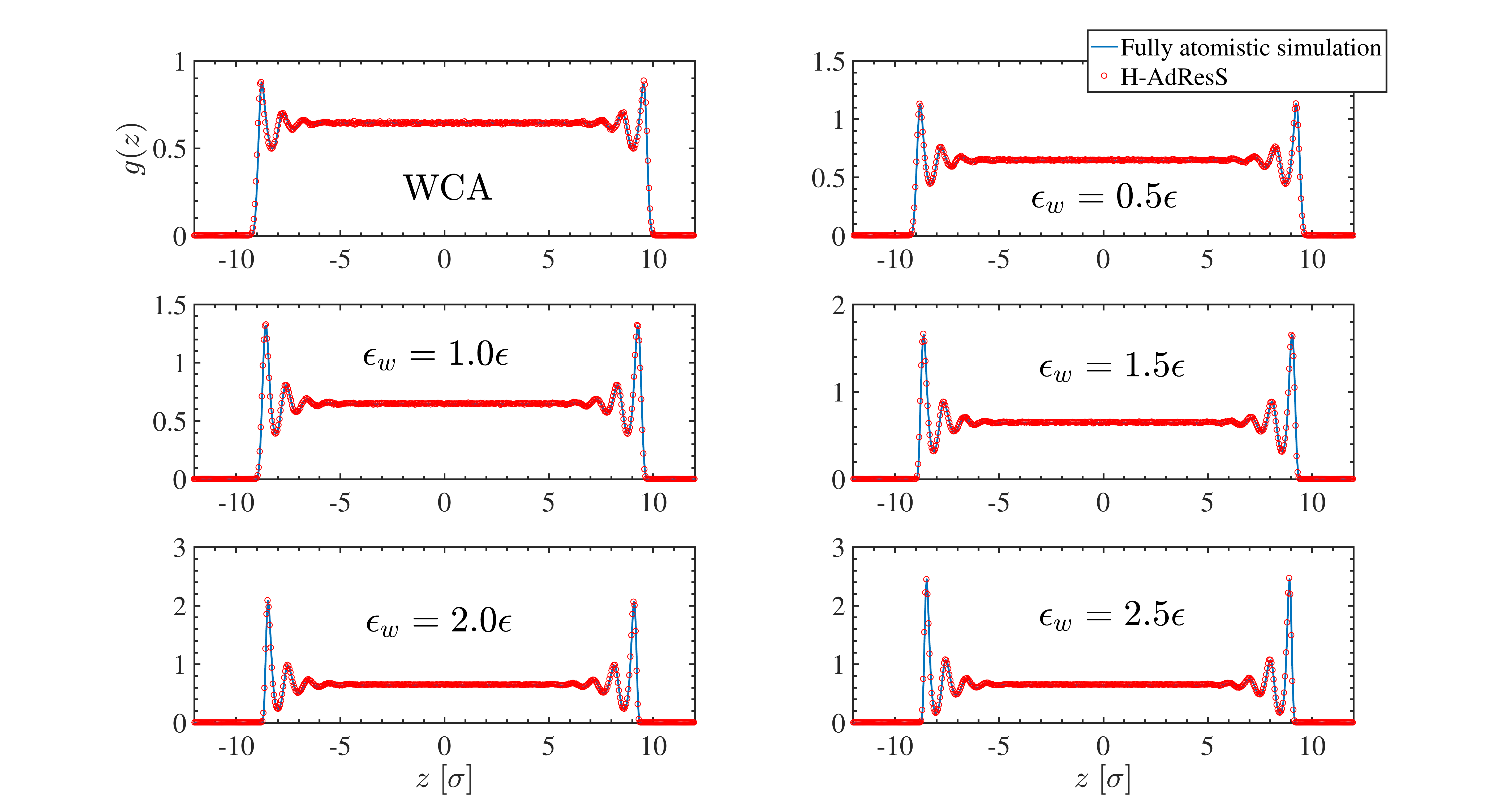}
 \caption{ Particle density profiles normal to the wall surface obtained for the LJ fluid with different wall interaction. The results are presented for fully atomistic simulation and {\tt H-AdResS}.} 
  \label{Fig:Eq_Dist_Single}
 \end{center}
 \end{figure*}
To investigate the structure of the liquid inside the AT region, we calculate the equilibrium particle distribution function normal to the wall, $g(z)$, and compare it with the results obtained from the fully atomistic simulation. The distribution function is obtained by binning the atomistic region along the z-axis and time averaging over all bins with equal $z$-coordinate. As it is shown in Figure \ref{Fig:Eq_Dist_Single}, the surface-induced layering structure is well reproduced by the {\tt H-AdResS} simulation. Moreover, the comparison of the bulk particle distribution ($g(z)$; for $\left|z\right|<5\sigma$) shows that the density of the atomistic region in the adaptive setup has converged to the atomistic reference density. These results confirm the suitability of our method to study confined liquids under the conditions here specified. 

At this point, it is important to mention that the identification of $\Delta H^{\rm s}$ with the chemical potential of the solvent is not straightforward in the case of a confined liquid. To compute $\Delta H^{\rm s}$ we use as a reference density the nearly constant value obtained in the central region between the parallel plates, whereas a full identification of $\Delta H^{\rm s}$ with the chemical potential should explicitly include the dependence with the distance normal to the surface. Hence, in the next section, to unambiguously enforce a constant chemical potential in an open system we introduce and validate the particle insertion algorithm for a bulk liquid system. 
 
\subsection{Particle insertion algorithm}\label{Sec:Particle_Excha}

The proposed grand canonical molecular dynamics method consist of two parts: first, the AT/IG constant chemical potential coupling that has been already discussed in the previous section as well as in Ref. \cite{SPARTIAN}. Second, we allow particle exchange between the IG region and an ideal gas reservoir used to control the chemical potential of the system. The details of the particle insertion algorithm applied on the IG region are the subject of this section.

We start by assuming that the IG region is in the grand canonical ensemble. The probability that the IG region, at temperature $T$, volume $V_{0}$ and chemical potential $\mu$, has exactly $N$ particles follow the Poisson distribution \cite{Landau-Lifshitz}

\begin{equation}
P(N) = \frac{(N^{*})^{N}\text{e}^{-N^{*}}}{N!}\, ,
\end{equation}

with $N^{*}$ the mean number of particles in the volume $V_{0}$. In the ideal gas case, $N^{*}$ can be written in terms of the chemical potential of the system

\begin{equation}
    N^{*} = \frac{V_{0}\text{e}^{\beta \mu}}{\lambda^3}\, ,
\end{equation}

with $\beta=1/k_{\rm B}T$ and $\lambda$ the mean thermal wavelength. In the limit $N,N^{*}\gg 1$ with $|(N-N^{*})/N^{*}|\ll 1$ we obtain

\begin{equation}
    P(N) = \frac{\text{e}^{-(N - N^{*})^{2}/2N^{*}}}{\sqrt{2\pi N^{*}}}\, .
\end{equation}

This is a normal distribution with mean value $N^{*}$ and standard deviation $\sqrt{N^{*}}$. In physical terms, this corresponds to the well known result for the isothermal compressibility $\kappa$ of the ideal gas, i.\ e.\ $\kappa = 1/\rho k_{\rm B }T$ with $\rho^{*} = N^{*} / V_{0}$. We are interested in fluctuations around a target density $\rho^{*}$, therefore, we rewrite $P(N)$ in terms of $\rho=N/V_{0}$ as

\begin{equation}
P(\rho) \sim \text{e}^{-k_{\mu}(\rho - \rho^{*})^{2}/2}\, ,
\end{equation}

where in principle $k_{\mu}=V_{0}/\rho^{*}$ but in the following we treat it as a free parameter related to the width of the distribution of possible values for the target density $\rho^{*}$. With this probability distribution, we introduce the Metropolis algorithm used for particle insertion. The probability to accept a move, namely, that the present density $\rho$ increases by $\nu$, is given by

\begin{equation}
    \text{acc}(\rho \to \rho+\nu)=\text{min}[1,\text{exp}(-k_{\mu}\nu(\nu + 2(\rho-\rho^{*})))]\, ,
\end{equation}

and correspondingly

\begin{equation}
    \text{acc}(\rho  \to \rho-\nu)=\text{min}[1,\text{exp}(-k_{\mu}\nu(\nu - 2(\rho-\rho^{*})))]\, .
\end{equation}

Concerning the fluctuations around the target density, we select values of $\nu$ according to the normal distribution

\begin{eqnarray}\label{eq:Prob_dist_ParEx}
P(\nu) = \sqrt{\frac{k_\mu}{2\pi}}\text{e}^{-\frac{k_\mu\nu^2}{2}}\, .
\end{eqnarray}

Finally, the particle insertion Monte Carlo moves are performed every $\Delta T_{exch}$ time steps during which the number of particles is averaged.

  \begin{figure}[h]
 \begin{center}
  \includegraphics[width=0.45\textwidth,angle=0]{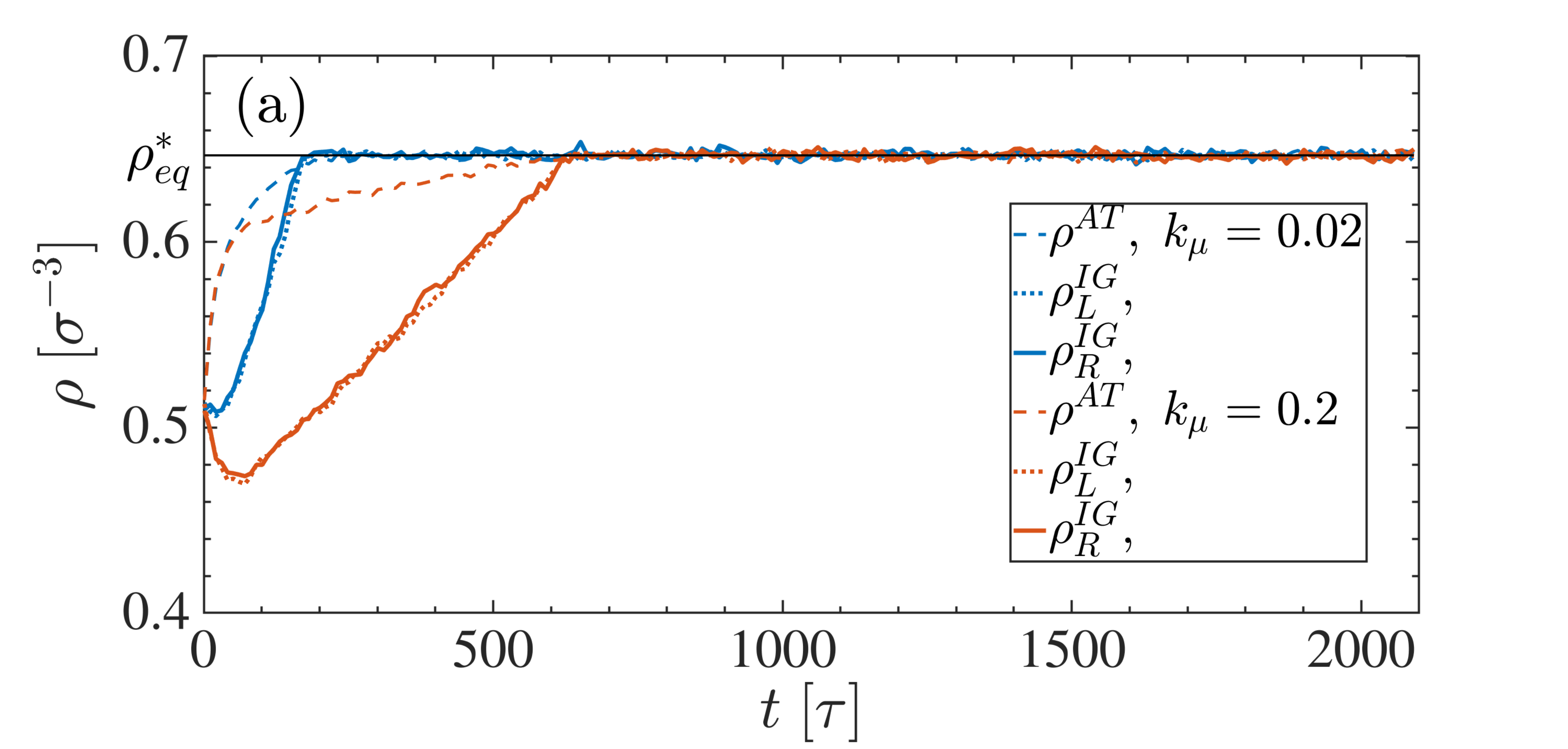}\\
  \includegraphics[width=0.45\textwidth,angle=0]{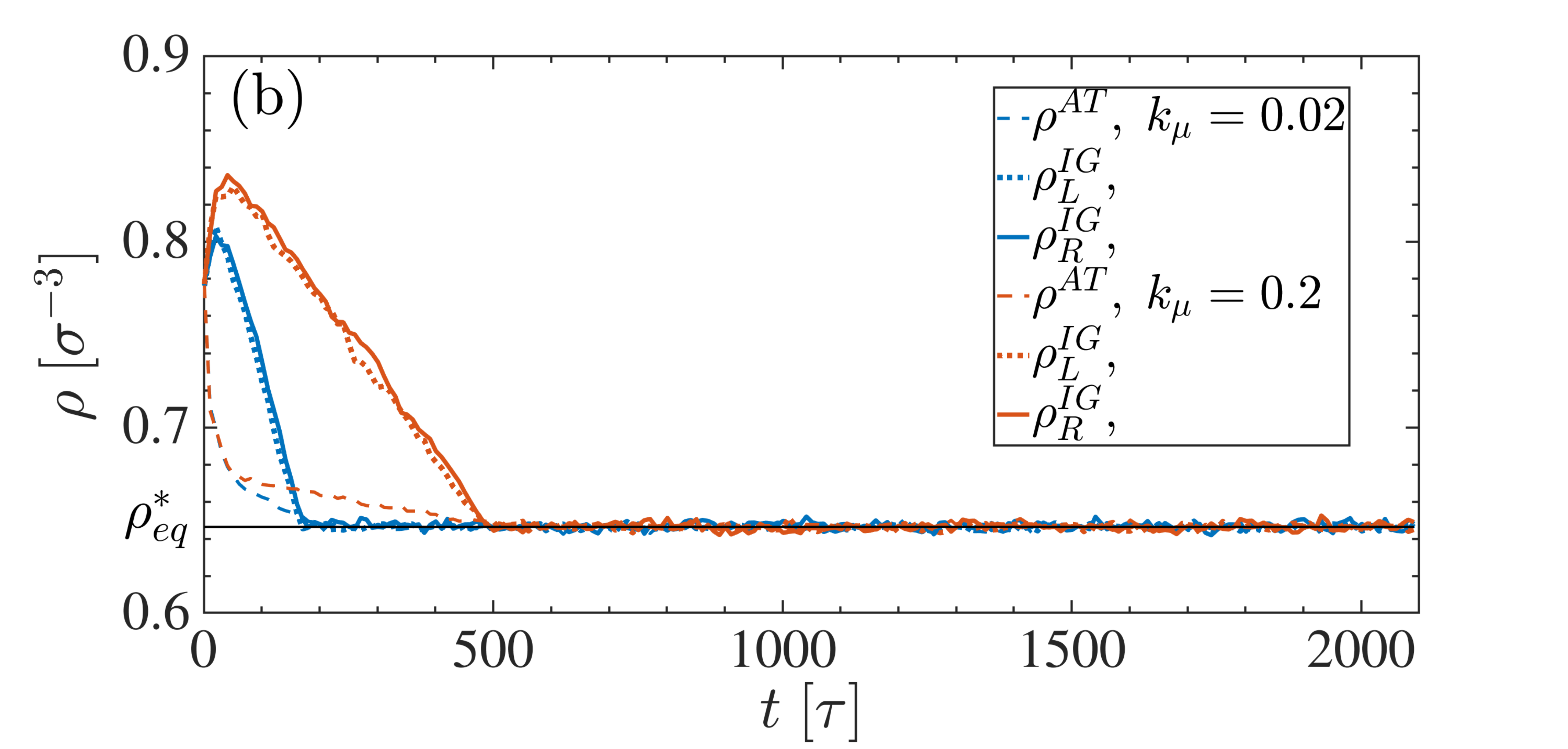}
 \caption{Evolution of the density in the left ($\rho_L^{IG}$), right ($\rho_R^{IG}$) ideal gas regions and atomistic ($\rho^{AT}$) regions when the system has initially lower (a) or higher (b) density than the equilibrium reference density ($\rho^{*}_{\rm eq}$).}
  \label{Fig:grand_Canonical}
 \end{center}
 \end{figure}

In conventional grand canonical simulations, the target system can interchange particles with an infinite ideal gas reservoir at constant chemical potential \cite{Frenkel-Smit}. Alternatively, the atomistic/coarse-grained coupling present in adaptive resolution simulations provides a suitable playground to sample the grand canonical ensemble \cite{Mukherji2013,DelleSite-Grand,Agarwal_2015,dellesite_GC2019}. In our particular case, the size of the reservoir can be substantially increased since particles in the IG region are non-interacting \cite{kreis_EPJST2015,SPARTIAN}. Alternatively, by introducing density fluctuations in the IG region, we also ensure interchange of particles with an infinite reservoir of ideal gas particles at constant chemical potential. 
	
To verify grand canonical conditions, we run an equilibrium simulation for a bulk LJ liquid, i.\ e.\ with the confining walls removed, at a given pressure $P=2.65\, \epsilon\sigma^{-3}$ and temperature $k_BT=2\, \epsilon$. These conditions define the target density $\rho_{\rm eq}^*=0.647\, \sigma^{-3}$ and the corresponding chemical potential. Upon obtaining the equilibrated all-atom configuration, "ghost" particles are placed in each reservoir and then {\tt H-AdResS} is performed using the Hamiltonian of Eq.\ \ref{eq:ham} and the obtained FEC terms corresponding to the target density. The velocity and force of the ghost particles are set to zero so they do not move during the {\tt H-AdResS} parameterization runs. The {\it on the fly} calculation of compensation of the drift and thermodynamic forces are updated every $5\times10^4$ time steps, during $10\times10^6$ time steps (200 iterations). The resolution interval is divided into 200 bins of size $\Delta \lambda=0.005$ and the length of the simulation box is uniformly discretised into slabs of size $\Delta x= 0.5\, \sigma$. We employed values of $c=1\, \epsilon$, $\sigma_\rho=2\, \sigma$\ and $\l = 4\, \sigma$ for smoothing and scaling the thermodynamic force. All simulations are performed with the {\tt LAMMPS} simulation package \cite{Plimpton19951,Heidari2016}, where the method is implemented.

To verify that the particle insertion protocol drives the system to a target density, $\rho^{*}$, we start with two versions of the system at the same temperature, but one at lower, $\rho<\rho^*$, and one at higher, $\rho >\rho^*$, density \cite{Boinepalli2003,eslami2007}. In both cases, we apply the FEC obtained from the target system to set the target chemical potential, and run the open boundary simulation using $\rho^{*}$ as an input for the particle insertion protocol. In Figure \ref{Fig:grand_Canonical} the evolution of density as a function of simulation time is presented in both cases, $\rho^{\rm IG}_{\rm L,R} < \rho^{*}$ (a) and $\rho^{\rm IG}_{\rm L,R} > \rho^{*}$ (b).

It is apparent from Figure \ref{Fig:grand_Canonical} that the density in the three regions, left, right and AT converges to the reference density in all cases, independently of the choice of insertion frequency $k_{\mu}$. In general this result verifies that the open simulation setup described corresponds to a constant chemical potential molecular dynamics simulation. The behaviour at short times indicates that the FEC works to restore the density in the atomistic region (dashed lines) by depleting (a) / increasing (b) the number of particles in the reservoir. The effect of the particle insertion algorithm is thus to bring the density of the reservoir (solid and dotted lines) to the target value and equate the chemical potential across different simulation regions.  Finally, the behaviour of the system as a function of $k_{\mu}$ is consistent with the interpretation given in terms of the width of the distribution of $\rho^{*}$. 

In the final section, we return to the confined liquid problem and use the particle insertion algorithm in a nonequilibrium molecular dynamics setup to induce a density gradient throughout the system.

\section{Induced density gradient and plane Poiseuille flow}\label{poiseuille} 

In this section, we start with the equilibrated configurations for the confined liquid considered in the previous section. Here, we use the Langevin thermostat to control the temperature in all regions at $k_BT=2\, \epsilon$ with damping coefficient $10\, \tau$. The time step is $\delta t=10^{-3}\, \tau$ and for each case, the total simulation run is $14\times 10^6$ time steps. To induce the density gradient between the atomistic region and ideal gas reservoirs, the particle exchange algorithm is applied independently in each reservoir. The reference number densities of particles in the left reservoir is set as the equilibrium number density $\rho_{L}^*=\rho_{eq}^*$. The reference number density in the right reservoir is increased  with respect to the equilibrium number density. We use $k_\mu=0.1\, \epsilon$ and $\Delta T_{exch}=100\, \delta t$ for all simulation sets. The reservoir (frozen) particles are also exchanged between the two reservoirs every $100$ time steps to balance the number of particles in the reservoirs. As both reservoirs are separated by repulsive walls, the number density of each reservoir is calculated over a smaller control volume than $V_{IG}$ which does not contain the depletion layer close to the right and left repulsive walls. Additionally, the width of the depletion layers changes when the solvent changes from LJ fluid (fully atomistic simulation) to ideal gas particles ({\tt H-AdResS}) which changes the total available volume. This difference in volume can be compensated by increasing the hard core size of the repulsive LJ potential between the walls and ideal gas particle such that the number density far from the depletion zone equals the one in fully atomistic simulations. In our study, however, such difference is negligible.

\begin{figure}[h]
 \begin{center}
     \includegraphics[width=0.5\textwidth,angle=0]{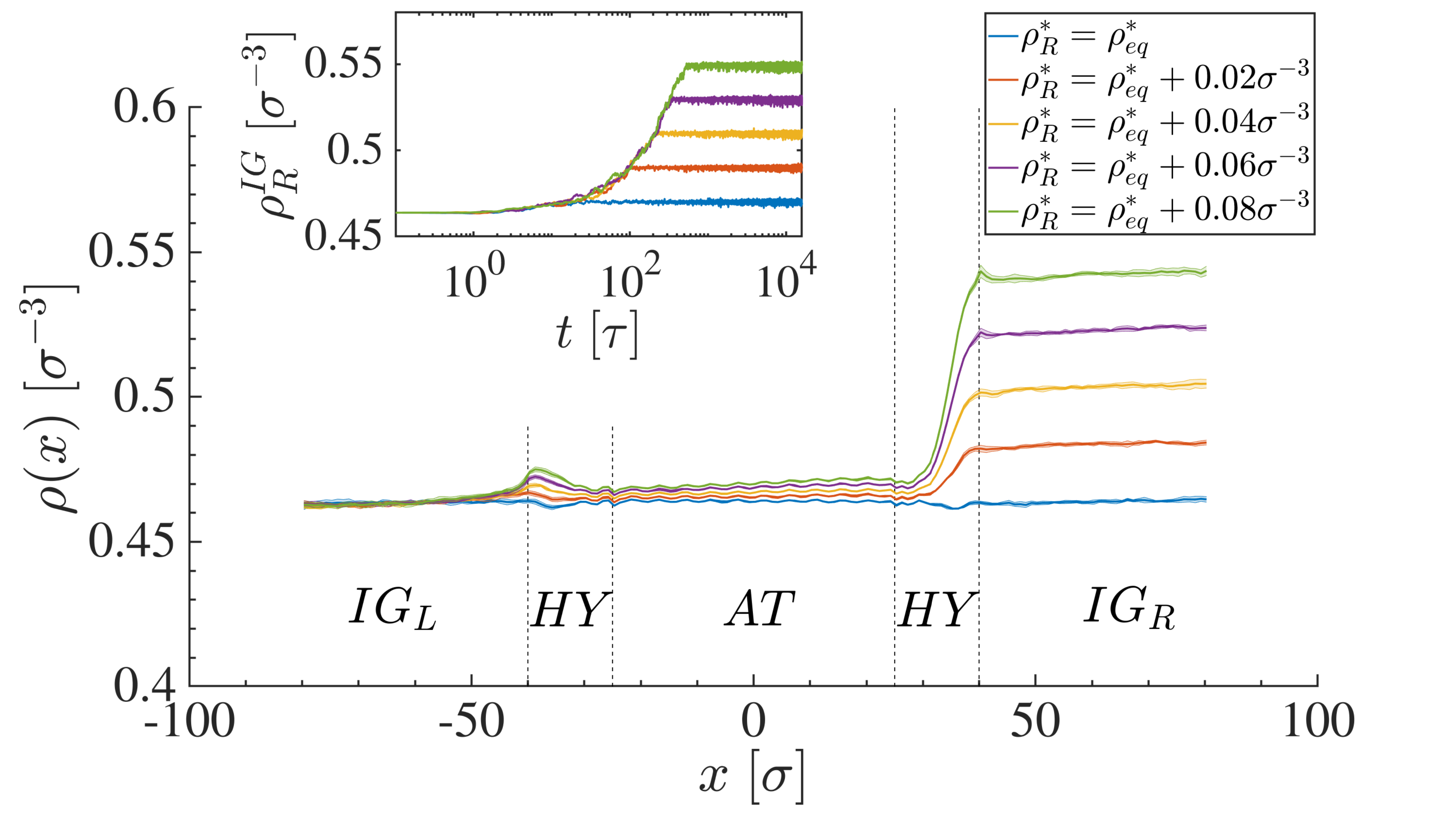}
  \caption{ Number density profiles shown for different reference values $\rho^*_{R}$. In all cases, the reference number density in the left reservoir is set to the equilibrium reference number density $\rho_{eq}^*$. Inset shows the time evolution of the number density in the right reservoir. } \label{Fig:Num_Density_Single}
 \end{center}
 \end{figure}

Concerning statistics, all values are reported using the block averaging method. For each case, we divide the total simulation run into seven uniform blocks of $2\times10^6$ time steps each. To remove the effect of the transient behavior, we do not consider the results of the first block (See inset of Figure \ref{Fig:Num_Density_Single}). The average values of each block are calculated and then we report the average and standard deviation values of all blocks. 

Figure \ref{Fig:Num_Density_Single} shows the number density profile of particles when the system reaches the steady state (as it is shown in the inset). It is apparent the induced density gradient in the atomistic region and the ripples observed there are generated by the interaction with the surface. As expected, the difference in densities between the right and left reservoirs is equal to the actual nominal difference. The density profile at the IG$_{L}$/HY interface exhibits bumps that become more distinct upon increasing the density gradient. These can be attributed to a mismatch in mobility due to particles changing identity from atomistic to non-interacting.  Thus, particles accumulate before entering the left reservoir, and once there the particle insertion algorithm flattens the profile to reach the reference density.   

 \begin{figure}[h]
 \begin{center}
\includegraphics[width=0.45\textwidth,angle=0]{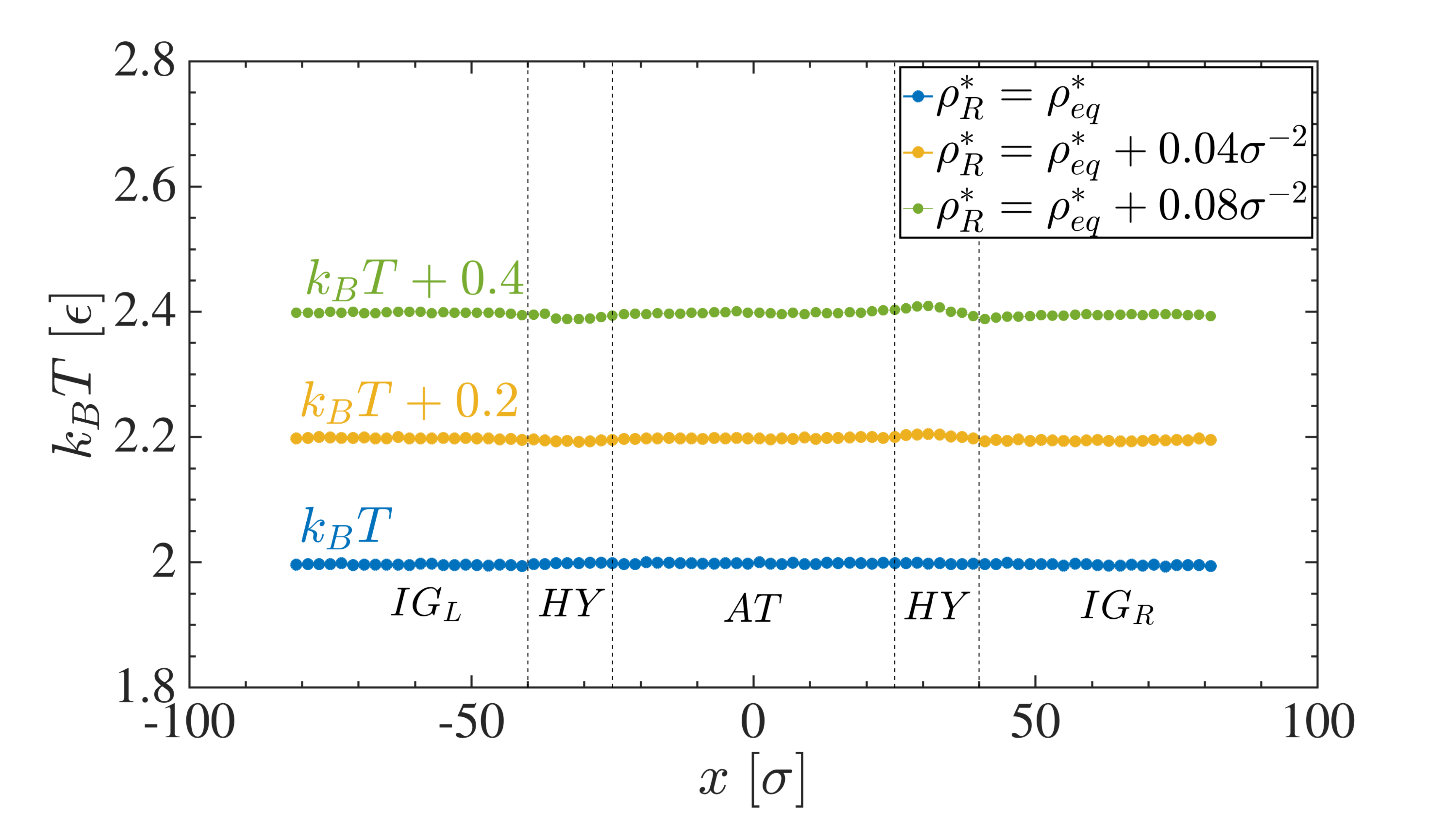}
  \caption{Temperature profiles of the system when the fluid-wall interaction strength is $\epsilon_w=2\, \epsilon$. For clarity, the profiles have been shifted by a constant quantity.} \label{Fig:Temp_Profile}
 \end{center}
 \end{figure}

To effectively isolate the effect of the induced density gradient in the atomistic region, it is necessary to guarantee that the temperature is constant throughout the simulation box. This is presented in Figure \ref{Fig:Temp_Profile}, clearly indicating a uniform temperature of the entire system. As a matter of fact, the maximum deviation of temperature from the mean value occurs in the case of induction of largest density gradient and it is $0.015\, \epsilon$ (less than 1 percent of the mean temperature). This is found to be in the middle of the hybrid region where the resolution of the ideal gas changes effectively into LJ fluid.

At this point, we are in the position to compare our simulation results with hydrodynamics, namely, the Poiseuille flow equation \cite{STALTER2018198}. A simple dimensional analysis justifies such a comparison. To this end, we emply the Knudsen number, defined as $\text{Kn} = \lambda/D_h$ with $\lambda=1/\sqrt{2}\rho\pi\sigma^{2}$ the mean free path of the fluid particles and $D_h=20\sigma$ the height of the channel. Knudsen numbers for the system under consideration vary between $\text{Kn}=0.014,\cdots,0.016$ indicating that a parallel can be drawn between continuum fluid dynamics and our simulation results \cite{Dong_AdvMechEng2012}.

A plane Poiseuille flow is created in a fluid with density $\rho$ and dynamic viscosity $\mu$ confined between infinitely long parallel plates, separated by a distance $D_h$, when a constant pressure gradient is applied along the axis of the channel $x$. The resulting flow is unidirectional. At low Reynolds number, the Navier-Stokes equation can be written as

\begin{equation}
    \frac{d^{2}u_{x}(z)}{dz^{2}} = \frac{1}{\mu}\frac{\Delta p}{\Delta x}\, ,
\end{equation}

where the axial velocity $u_{x}(z)$ is only a function of the $z$-coordinate and the applied pressure gradient $\Delta p/\Delta x$ is constant. Using the boundary conditions $u_{x}(z=\pm D_h/2)=0$ (no-slip condition), we obtain

\begin{equation}\label{eq:ux}
    u_{x}(z) = \frac{1}{\mu}\frac{\Delta p}{\Delta x} \left[ z^{2} - \frac{D_h^{2}}{4}\right]\, .
\end{equation}

\begin{figure}[h]
 \begin{center}
\includegraphics[width=0.45\textwidth,angle=0]{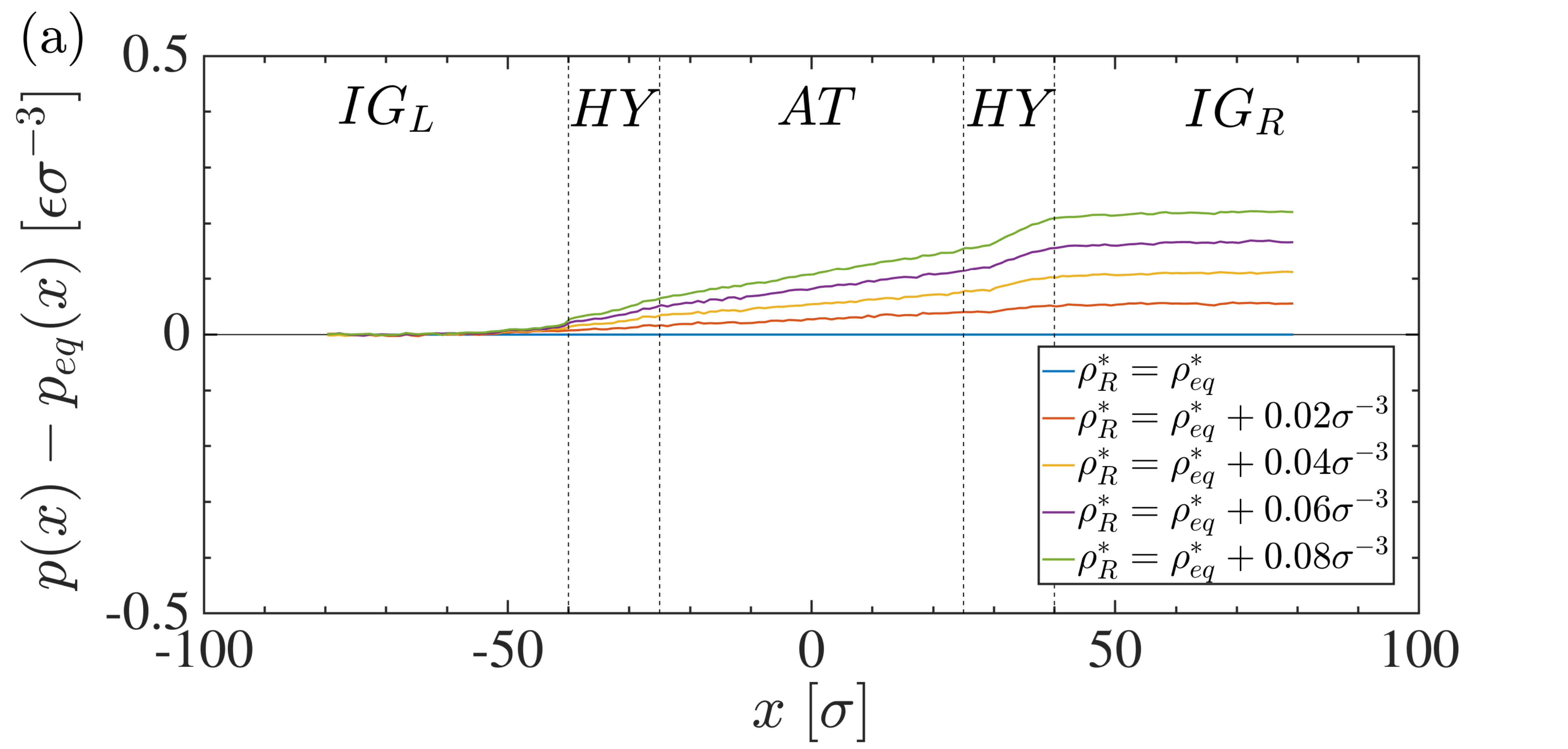}
\includegraphics[width=0.45\textwidth,angle=0]{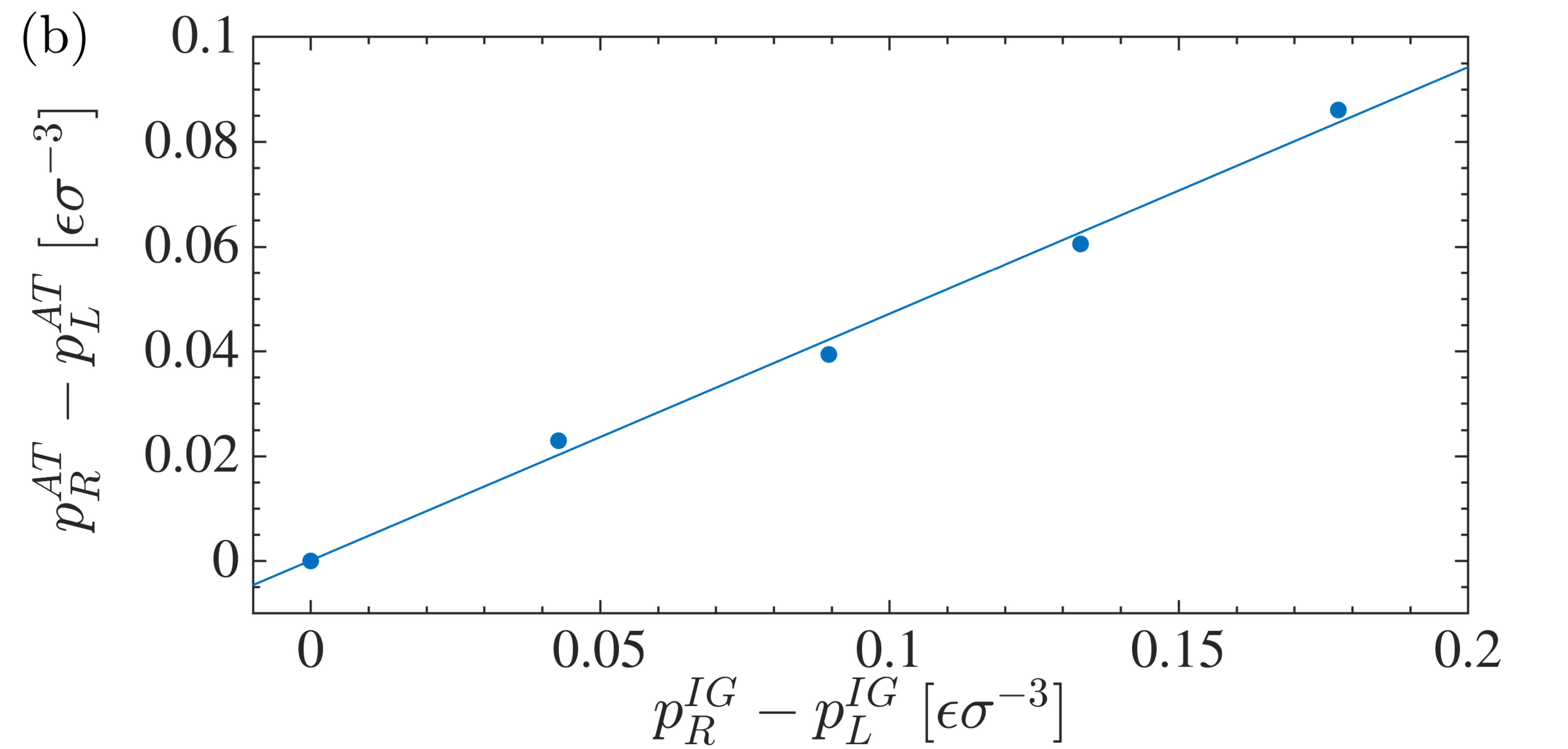}
  \caption{(a) Difference of pressure profiles of the systems with respect to the equilibrium pressure profile $p(x)-p_{eq}(x)$. 
  (b) Comparison between pressure difference of reservoirs $p_{R}^{IG}-p_{L}^{IG}$ and the pressure difference calculated at the boundary of atomistic region $p_{R}^{AT}-p_{L}^{AT}$. } \label{Fig:Press_Profile}
 \end{center}
 \end{figure}

Before comparing this result with the velocity profile obtained from simulations we emphasise two aspects. First, to obtain the result in Eq.\ \eqref{eq:ux} one assumes that the fluid is incompressible, and the LJ system under consideration is not. However, the Navier-Stokes equation for a compressible fluid contains a term proportional to $\nabla \cdot \mathbf{u}$ that in our case is equal to zero because the flow is unidirectional along the $x$-direction and the magnitude of the flow velocity does not change along this axis. Second, the result in Eq.\ \eqref{eq:ux} was obtained by assuming that $\Delta p/\Delta x$ is constant. In our model we need to verify that the constant pressure gradient induced in between the right and left reservoirs creates a constant pressure gradient across the AT region. Plot of pressure profiles (Figure \ref{Fig:Press_Profile} (a)) for different induced densities, in the case of purely repulsive interaction with the walls, indicates that this is indeed the case.  

 As a matter of fact, we find that there is a linear relation between the pressure difference measured across the AT region, $p^{\rm AT}_{\rm R}-p^{\rm AT}_{\rm L}$ and the nominal pressure difference $p^{\rm IG}_{\rm R}-p^{\rm IG}_{\rm L}$ for different values of the induced density, as indicated in Fig.\ \ref{Fig:Press_Profile} (b). The slope of the line is 0.47.  This value coincides with the constant pressure gradient across the resolution, namely $\frac{L_{AT}}{L_{AT}+2L_{HY}+L_{corr}}=\frac{5}{10}$. The quantity $L_{Corr}=20\, \sigma$ measures the distance from the interface of the HY region to the left reservoir up to the point at which the pressure reaches the equilibrium pressure. The constant pressure gradient obtained in the AT region can be understood in terms of the virial expansion of the pressure in terms of the density for a LJ system, namely, relatively small gradients in density (of maximum $0.08\, \sigma^{-4}$ here) induce ideal-gas-like response in the atomistic system. 
 
  \begin{figure*}[h]
 \begin{center}\includegraphics[width=1.0\textwidth,angle=0]{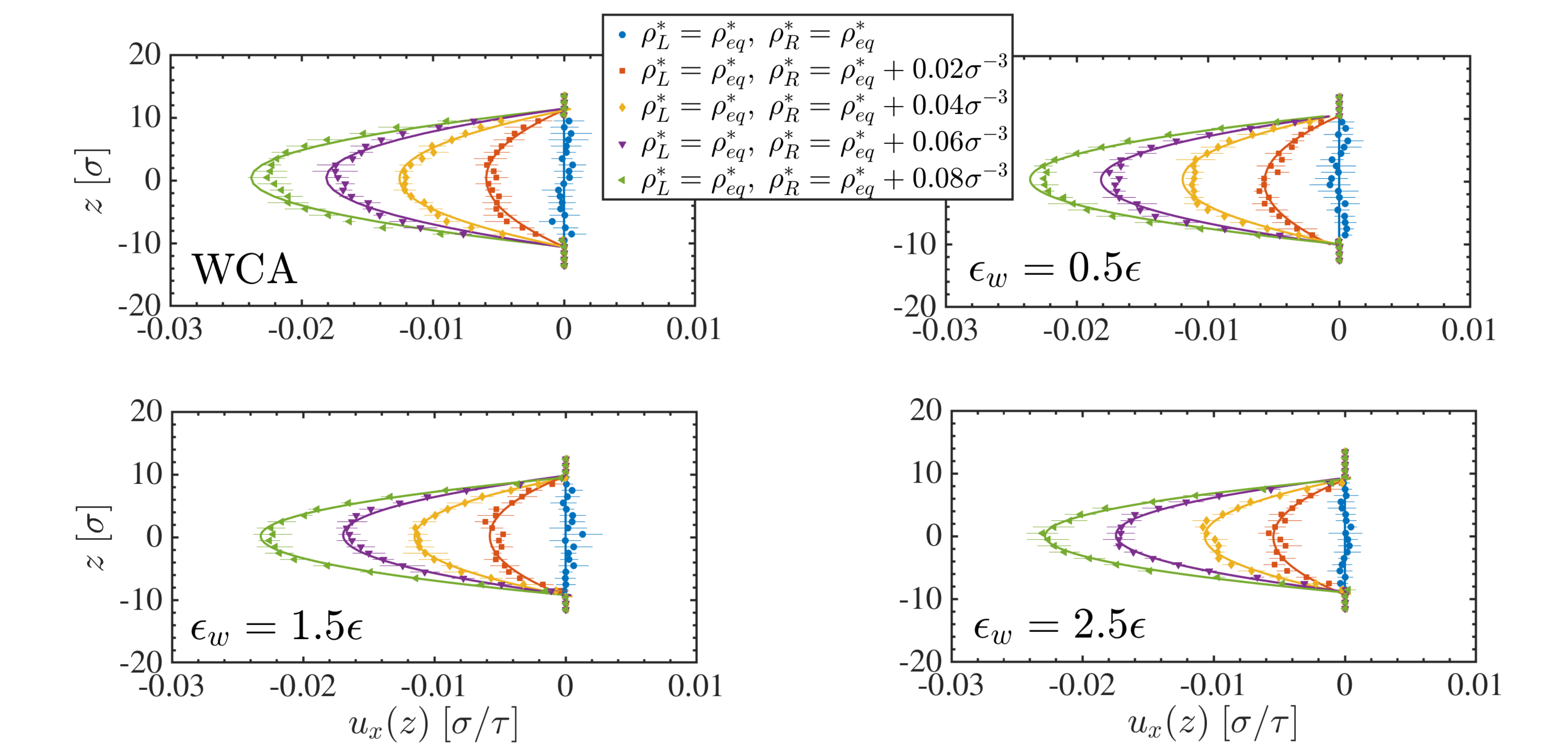}
  \caption{ Velocity profiles of the pressure driven flow are shown for different density gradients induced by controlling the density at the right reservoir ($\rho_R^*$). The density at the left reservoir is set to equilibrium density $\rho_L^*=\rho^{eq}$. The solid lines are the parabolic fitting to each set of data points.} \label{Fig:Velo_Profile}
 \end{center}
 \end{figure*}
 
To compare with the result expected from Eq.\ \eqref{eq:ux}, we calculate the corresponding average of $u_{x}(z)$ in the atomistic region and obtain the velocity profiles plotted in Figure \ref{Fig:Velo_Profile} for all liquid-wall interactions. The solid lines represent the parabolic functions fitted to the data points. In all cases and apart from relatively small statistic fluctuations, the simulated velocities follow closely the expected parabolic profile. 

 \begin{figure}[h]
 \begin{center}
\includegraphics[width=0.45\textwidth,angle=0]{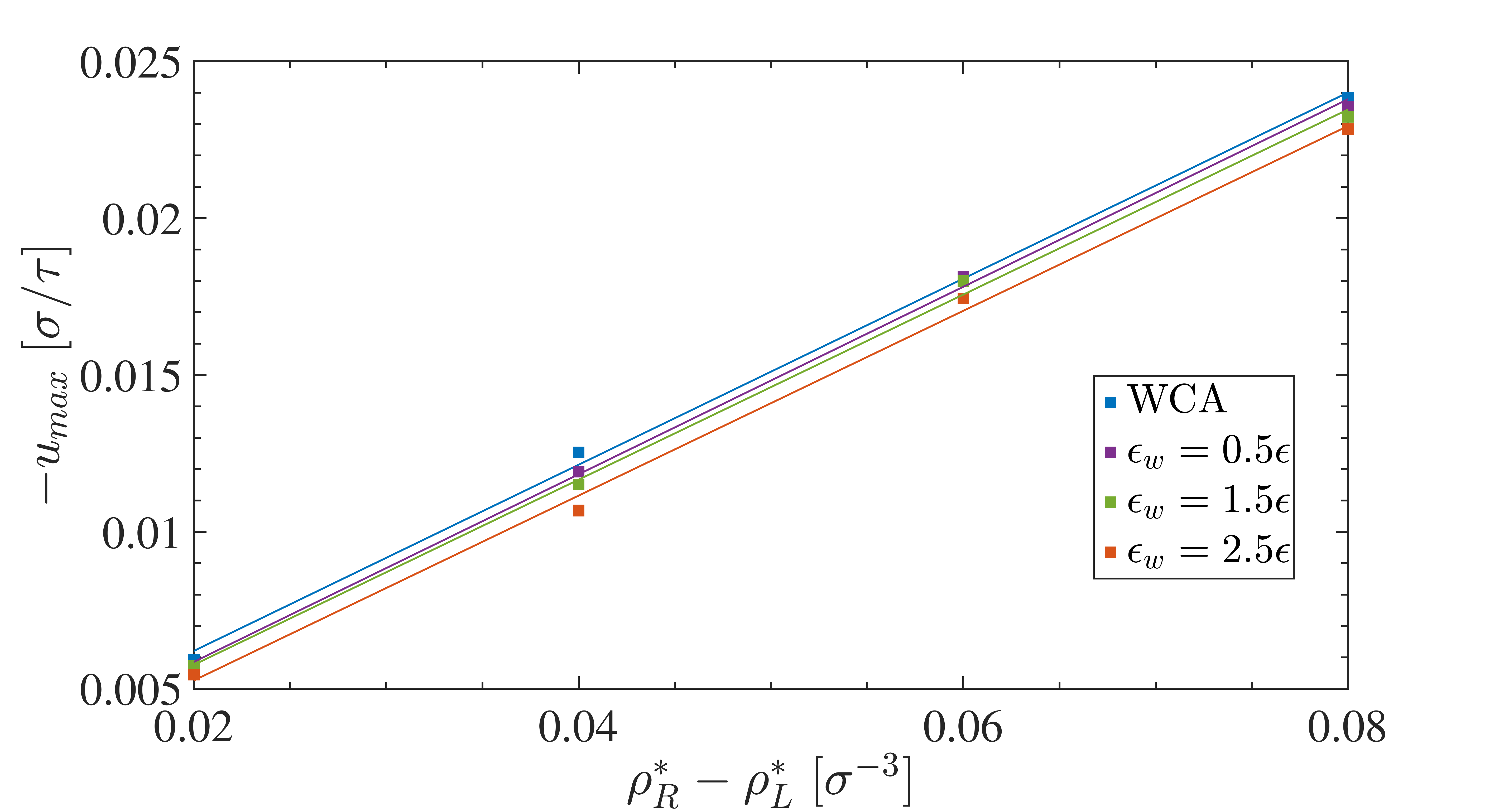}
  \caption{ Maximum velocity of the Poiseuille flow vs. different relative densities between the reservoirs  ($\rho_R^*-\rho_L^*$). The density at the left reservoir is set to equilibrium density $\rho_L^*=\rho^{eq}$. The solid lines are linear fits to the data points.} \label{Fig:Velx_vs_rho}
 \end{center}
 \end{figure}

The resulting velocity profile implies that the maximum velocity occurs at the center of the channel $-u_{max}=u_{x}(z=0)$. This maximum velocity is linearly proportional to the nominal density difference, i.\ e.\ $u_{max}\propto (\rho^*_{R}-\rho^*_L)$, as presented in Figure \ref{Fig:Velx_vs_rho}.

 \section{Discussion and Conclusions}\label{conclusions}

In computer simulations, the investigation of a large closed system allows one to sample the grand canonical ensemble. A subdomain of volume $V$ and an average number of particles $\langle N\rangle$ inside a system with a fixed number of particles $N_{0}$ and volume $V_{0}$ is considered to be in the grand canonical ensemble provided the condition $V/V_{0}\approx 0.01$ holds \cite{Heidari2018Frenkel}. For the Lennard-Jones systems investigated here, a subdomain of volume $V\sim (10\, \sigma)^{3}$, with the estimate for the correlation length of the system as $10\, \sigma$, would require $V_{0}\sim 100 V$. With a density $\rho = 0.647\, \sigma^{-3} $, the simulation would need $N_{0}\sim 10^{6}$. This number is one order of magnitude larger than the number of particles considered here, which already highlights the advantage of using the proposed method. Additionally, we emphasise that a detailed assessment of the relative atomistic, hybrid and ideal gas region sizes might suggest that we could further reduce the size of the system without affecting our main results. 

Instead of the direct approach discussed above, it is perhaps more convenient to use one of the available grand canonical molecular dynamics methods \cite{Frenkel-Smit,adams1975,Cagin-Pettitt-MolSim6-5-1991,Papadopoulou-etal-JCP98-4897-1993,Pettitt1997,Thompson-etal-JCP109-6406-1998,Shroll1999,Boinepalli2003,eslami2007,Boda2008}. A common ingredient in most approaches consists of inserting particles with a given system-dependent probability. In general, when the system under consideration is simple in terms of the force field used for its description, or if the system is at low density/concentration conditions, this is the method of choice.  Far away from such conditions, the particle insertion protocol becomes highly inefficient. In this respect, the adaptive resolution framework constitutes an alternative for existing methods. In particular, for the Hamiltonian adaptive resolution discussed in Ref. \cite{SPARTIAN}, the target system is in contact with a reservoir of ideal gas particles at constant temperature, volume and, by ensuring a uniform density across the simulation box, also at constant chemical potential.  The combination with a particle insertion algorithm operating in the ideal gas region guarantees a reservoir of infinite size, thus completing the definition of grand canonical ensemble.  Therefore, the method proposed here is a robust strategy to perform open-boundary molecular dynamics simulations, mainly when the system under consideration is dense, or highly concentrated in the case of mixtures, and regardless of the complexity introduced by the force field description.

A straightforward change in the geometry and periodic boundary conditions in the simulation box allows one to decouple the ideal gas reservoir. Hence, it is possible to simultaneously impose different temperature, density and concentration conditions on the system. In the particular case of induced density gradients \cite{Todd-Daivis,Dong_AdvMechEng2012,Hannon-PLA119-174-1986,Koplik-etal-PhysFluidsA1-781-1989,Heinbuch-Fischer-PhysRevA40-1144-1989}, current non-equilibrium molecular dynamics methods introduce external forces that might introduce artefacts in the simulation resulting from non-conservation of linear momentum. Instead, the method proposed here conserves momentum on average. Moreover, the simplicity of the reservoir gives the possibility to study different out-of-equilibrium conditions for complex molecular systems, which constitutes a significant improvement over state-of-the-art simulation methods. 

To conclude, in this work, we presented a method to perform open-boundary molecular dynamics simulations. We used the {\tt H-AdResS} framework where the atomistic target system is in physical contact with a reservoir of non-interacting particles at constant temperature, volume and chemical potential. In addition to the straightforward calculation of the chemical potential, the use of  {\tt H-AdResS} allows one to study liquid mixtures directly. In this context, we introduced a particle insertion/deletion algorithm that operates, at minimal computational expenses, on the ideal gas reservoir. Approaches exploiting similar ideas are available in the literature \cite{Cracknell-etal-PRL74-2463-1995,Cagin-Pettitt-MolSim6-5-1991}, however, they lack the flexibility provided by the coupling to the ideal gas system. The proposed method allows one to perform constant chemical potential simulations in various conditions. More importantly, by studying pressure-driven flow through a channel, we showed that it is also possible to perform well-controlled non-equilibrium molecular dynamics simulations. 

\section{Acknowledgments}

The authors thank Pietro Ballone and Dominic Spiller for a critical reading and insightful comments. M.H., K.K., R.P. and R.C.-H. gratefully acknowledge funding from SFB-TRR146 of the German Research Foundation (DFG). This work has been supported by the European Research Council under the European Union’s Seventh Framework Programme  (FP7/2007-2013)/ERC  Grant  Agreement No. 340906-MOLPROCOMP. This project received funding from the European Research Council (ERC) under the European Union's Horizon 2020 research and innovation program (Grant 758588).

 \bibliography{main}
\end{document}